\documentclass[12pt]{article}
\usepackage{eufrak}
\newcommand{\be}{\begin{equation}}
\newcommand{\ee}{\end{equation}}
\newcommand{\bea}{\begin{eqnarray}}
\newcommand{\eea}{\end{eqnarray}}

\def\km{{\mathfrak M}}

\title{``Dynamical'' non-minimal higher-spin interaction and gyromagnetic ratio $g=2$}
\author{I. Ots$^{a)}$, R.Saar$^{b)}$, R.-K. Loide$^{c)}$, H. Liivat$^{b)}$}
\date{}

\begin{document}

\maketitle
\noindent
\begin{center}
{\small {\it a) Institute of Physics, Tartu University, Riia 142, 51014 Tartu, Estonia} \\
{\it b) Institute of Theoretical Physics, Tartu University, T\"ahe 4, 51010 Tartu, Estonia} \\
{\it c) Department of Physics, Tallinn Technical University, Ehitajate 5, 19086 Tallinn, Estonia}}
\end{center}
\vspace{0.5cm}

\begin{abstract}
The field-dependent invariant representation 
(the ``dynamical'' representation) of the Poincar\'e algebra 
is considered as a dynamical principle in order to get the 
corresponding ``dynamical'' electromagnetic coupling for higher 
spins ($s\geq 1$). If in lower-spin ($s=0,1/2$) cases the 
``dynamical'' coupling is taken to coincide with the minimal 
electromagnetic coupling the higher-spin coupling is inevitably 
non-minimal, containing a term linear in the field 
strength tensor $F_{\mu\nu}$. This term leads to $g=2$.
\end{abstract}

\vspace{1cm}
\noindent
It is almost a common agreement that the value for the gyromagnetic ratio
$g$ for all ``truly elementary'' charged particles of any spin is $g=2$.
To justify this choice usually two main arguments are given \cite{ferrara,jackiw}:

1) $g=2$ value must hold to guarantee a good high-energy behaviour
of scattering amplitudes (It was first shown by Weinberg more than thirty 
years ago \cite{weinberg}).

2) In the case of $g=2$ the Bargmann-Michel-Telegdi \cite{bargmann} equation
of motion of the polarization vector takes its simplest form:
$$
\frac{dS_\mu}{d\tau} = \frac{e}{m} F_{\mu\nu}S^\nu.
$$

However, these reasons can be qualified as ``practical needs''.
Obviously the $g=2$ value lays on some fundamental theoretical
grounds. Such grounds have been looked for in recent years.
Ferrara, Porrati and Telegdi \cite{ferrara} have shown that the requirement 
that tree-level amplitudes should not violate unitarity up to 
C.M. energies $E \gg M/e$ (good high-energy behaviour) is 
equivalent to the requirement of a smooth $M\to 0$ fixed-charge 
limit. A symmetry principle that suggests the $g=2$ value
has been also offered by Jackiw \cite{jackiw}. He has pointed to the fact that
the nonelectromagnetic gauge invariance possessed by free 
Lagrange density for spin-$1$ fields is preserved when the fields 
couple to the external electromagnetic field, provided $g=2$.
In what follows we will offer another symmetry principle which
leads to the $g=2$ value for higher spins.

Since the gyromagnetic ratio is determined by the form 
of the electromagnetic coupling, the search for the symmetry 
principle that would generate the $g=2$ value is also the search
for the dynamical symmetry principle for building a consistent
higher-spin electromagnetic interaction theory. Building such a 
theory is an old and difficult problem. Since the 
sixties much work has been done to solve the problem, however, the 
theory is far from its completion. 

One thing seems to be sure: the coupling of higher-spin charged 
particles to the electromagnetic field cannot be minimal.
That the minimal electromagnetic coupling leads to inconsistencies 
was known more than fourty years ago (see, for example, the papers of 
Federbush \cite{federbush}, Johnson and Sudarshan \cite{johnson}, 
Arnowitt and Deser \cite{arnowitt}). At the end of sixties the 
troubles connected with the acausal propagations in the higher-spin 
interaction theories with the minimal coupling were demonstrated 
(though uncompletely \cite{cox}) by Velo and Zwanziger \cite{velo}.

It became clear that the only way to get the consistent higher spin 
theory is to introduce the non-minimal interaction into the theory. 
The question is how to find the right non-minimal coupling. 
In the spin-$1$ case, to overcome the 
bad high-energy behaviour of the scattering amplitudes, a suitable 
non-minimal coupling term linear in the $F_{\mu\nu}$ was added
in fact by hand. From this term also the $g=2$ value arises.

Various possibilities with non-minimal $F_{\mu\nu}$ terms
and their consequences in spin-$3/2$ theories have also been 
analyzed. Recently, the general non-minimally coupled massive 
spin-$3/2$ models have been considered by Deser, Pascalutsa, and 
Waldron \cite{deser}. However, in that brilliant paper the model has 
been viewed as an effective theory and the question of the 
possible dynamical symmetry principle has been out of scope of 
their investigations.

Recently, we have proposed a symmetry principle that determines
the non-minimal coupling to the special (laser) field for a
charged elementary particle of any spin \cite{saar}. We have considered
the field-dependent invariant representation (the ``dynamical''
representation) of the Poincar\'e algebra as a dynamical
principle that determines the ``dynamical'' electromagnetic
coupling. The ``dynamical'' coupling for arbitrary $s\geq 1$
spin contains a term linear in the field strength. As it appears, 
this term leads to $g=2$ for higher spins.

The ``dynamical'' representation of the Poincar\'e algebra for
lower spins ($s=0,1/2$) has first been used by Chakrabarti \cite{chakrabarti}
to investigate the possibilities of extending the Volkov exact 
solution cases \cite{volkov}. A general ``dynamical''
representation for an arbitrary spin has been constructed in \cite{saar}.
The representation is built up by introducing a special external
electromagnetic field into the free Poincar\'e algebra. The
``dynamical'' representation is constructed from the generators 
of the free Poincar\'e algebra and the external field in such 
a way that the new, field-dependent generators obey the 
commutation relations of the free Poincar\'e algebra. Now, 
analogously to the free-particle theory, the wave equations 
with respect to the ``dynamical'' representation can be 
constructed. These equations describe the ``dynamical''
interaction of the particles with the external field.
As it has been shown already by Chakrabarti \cite{chakrabarti}, the simplest
way to build the ``dynamical'' representation is to
introduce the external field by a nonsingular transformation
$U$. Consequently, the problem is to find a field-dependent
$U$, such that the transformed Poincar\'e generators
\bea
\pi_\mu = U P_\mu U^{-1} \nonumber \\
{\km}_{\mu\nu} = U M_{\mu\nu} U^{-1} , 
\label{eq1}
\eea
where $P_\mu = \mathrm{i} \partial_\mu$, 
$M_{\mu\nu} = L_{\mu\nu}+S_{\mu\nu}$ with 
$L_{\mu\nu} = x_\mu P_\nu - x_\nu P_\mu$ and 
$S_{\mu\nu}$ as the generators of the finite-dimensional
representation of the Lorentz group, would obey the 
commutation relations of the free-particle theory, i.e.
\bea
\left[\km_{\mu\nu},\km_{\rho\sigma}\right] &=& {\mathrm{i}} 
(g_{\mu\sigma}\km_{\nu\rho}+g_{\nu\rho}\km_{\mu\sigma}
-g_{\mu\rho}\km_{\nu\sigma}-g_{\nu\sigma}\km_{\mu\rho}), \nonumber \\
\left[\km_{\mu\nu},\pi_{\sigma}\right] &=& {\mathrm{i}}
(g_{\nu\sigma}\pi_{\mu}-g_{\mu\sigma}\pi_{\nu}),\nonumber \\
\left[\pi_{\mu},\pi_{\nu}\right] &=& 0 .
\label{eq2}
\eea

Such an operator $U$ can be found in an arbitrary spin case 
for a special ``plain wave'' field
\be
A_{\mu}=A_{\mu}(\xi),\: \xi = k\cdot x
\label{eq3}
\ee
with the Lorenz gauge
\be
\partial_{\mu}A^\mu = 0 .
\label{eq4}
\ee 

The arbitrariness of $U$ is avoided by specifying the transformation
in such a way that for spin-$0$ and spin-$1/2$ the ``dynamical''
interaction would coincide with the minimal coupling and the
transformed wave function 
would give the Volkov's exact solution. Then in the spin-$0$ case 
the ``dynamical'' representation is obtained by transforming the free
Poincar\'e generators by the unitary operator \cite{chakrabarti}
\be
U_{0} = exp\biggl\{\mathrm{i}\int\frac{d\xi}{2(k\cdot P)}
\left[ 2e\, P\cdot A(\xi) - e^2 A^2(\xi)\right]\biggr\}.
\label{eq5}
\ee 
In the case of spinning particles one has to multiply the 
operator $U_{0}$ by the spin term, i.e.
\be
U = U_{0}\cdot U(s)\,,
\label{eq6}
\ee
where for arbitrary spin $s$ the spin part of the transformation 
can be given as \cite{saar}
\be
U(s) = exp\biggl\{-\mathrm{i}\frac{e}{2(k\cdot P)} 
\left[ k_{\mu}A_{\nu}-k_{\nu}A_{\mu}\right] S^{\mu\nu}
\biggr\} .
\label{eq7}
\ee
In eqs. (\ref{eq5}) and (\ref{eq7}) $\:k\cdot P=k_{\mu}P^{\mu}$ 
is an operator which commutes with any other one. 
It plays a special role in the theory. The details connected 
with the inverse operator $(k\cdot P)^{-1}$, also the physical 
meaning and significance of the ``dynamical'' representation
one finds in \cite{chakrabarti,beers}.

By applying transformation (\ref{eq6}) to the Poincar\'e
operator $P_\mu$, one gets
\be
P_\mu \to UP_{\mu}U^{-1} =\pi_\mu = P_{\mu}+\frac{e}{2(k\cdot P)}
\,k_{\mu}(e A^2-2 A\cdot P - F_{\sigma\rho}S^{\sigma\rho}).
\label{eq8}
\ee
Now one is ready to transform a general free Klein-Gordon 
equation
\be
(P^2-m^2)\psi=0
\label{eq9}
\ee
into the equation in the ``dynamical'' representation
\bea
U (P^2 - m^2)\psi=U (P^2 - m^2)U^{-1}U\psi= \nonumber \\
(\pi^2 - m^2)\psi_d=(D^2-eF_{\mu\nu}S^{\mu\nu}-m^2)\psi_d=0,
\label{eq10}
\eea
where $D^2=D_{\sigma}D^\sigma$, $D_{\sigma}=P_{\sigma}-eA_{\sigma}$
and $\psi_d=U\psi$. Since $S^{\mu\nu}$ is the Lorentz generator, 
equation (\ref{eq10}) describes a spectrum of spins. However, 
for each of these there is the same spin-dependent non-minimal term
$eF_{\mu\nu}S^{\mu\nu}$ which suggests that the gyromagnetic ratio $g=2$.

In what follows the simplest and the most familiar
spin-$1$ case will be considered in more detail.
To get a one-spin theory one must eliminate all superfluous spins in the
Lorentz representation in the Klein-Gordon equation (\ref{eq9}). 
This can be achieved by putting subsidiary conditions to the 
equation. In the massive spin-$1$ case the subsidiary condition is 
already in the Proca equation
\be
\biggl\{(P^2-m^2)g_{\mu\nu}-P_{\mu}P_{\nu}\biggr\}\phi^\nu = 0,
\label{eq11}
\ee
which can equivalently be written as the equation with the
subsidiary condition
\bea
(P^2-m^2)\phi_{\mu}=0, \nonumber \\
P_{\nu}\phi^{\nu}=0.
\label{eq12}
\eea
By applying the $U$-transformation to these spin-$1$ free particle 
equations, one gets the equations in 
``dynamical'' representation:
\be
\biggl\{(D^2-m^2)g_{\mu\nu}-D_{\mu}D_{\nu}
-2{\mathrm{i}}eF_{\mu\nu}\biggr\}\phi^\nu_d = 0
\label{eq13}
\ee
and
\bea
(D^2-m^2)\phi_\mu - 2{\mathrm{i}}eF_{\mu\nu}\phi^\nu_d =0,\nonumber \\
D_\nu\phi^\nu_d=0.
\label{eq14}
\eea
For deduction equations (\ref{eq13}) and (\ref{eq14}) we have used 
the fact that for spin-$1$
\be
S_{\mu\nu}=-{\mathrm{i}}e_{\mu\nu}={\mathrm{i}}(E_{\mu\nu}-E_{\nu\mu}),
\label{eq15}
\ee
where $E_{\mu\nu}$ generate the Weyl basis of the set of $4\times 4$ 
matrices
\be
(E_{\mu\nu})^{\rho}_{\:\:\sigma}=g_{\mu}^{\:\:\,\rho}g_{\nu\sigma}.
\label{eq16}
\ee
Thus, for spin-$1$ one gets
\be
(F_{\mu\nu}S^{\mu\nu})^{\rho}_{\:\:\,\sigma}
=-{\mathrm{i}}(F_{\mu\nu}e^{\mu\nu})^{\rho}_{\:\:\,\sigma}=
2{\mathrm{i}}F^{\rho}_{\:\:\,\sigma}.
\label{eq17}
\ee
Equations (\ref{eq13}) and (\ref{eq14}) are well-known, 
describing the coupling of a charged spin-$1$ particle to the 
electromagnetic field. The non-minimal linear in the $F_{\mu\nu}$ 
term guarantees a good high-energy behaviour of the scattering amplitudes 
and leads to the $g=2$ value of a gyromagnetic ratio.

Quite often it is stated that the coupling in eqs. (\ref{eq13}) and (\ref{eq14})
is the minimal one, i.e. it can obtained by making the substitution
$P_\mu\to P_\mu-eA_\mu$ in the free Proca equation (\ref{eq9}).
Indeed, since the procedure $P_\mu\to D_\mu$ is not unique in (\ref{eq11})
one can use a trick here \cite{aitchison}:
\be
-P_\mu P_\nu\to  -P_\mu P_\nu +k\left[ P_\mu, P_\nu\right]\to 
-D_\mu D_\nu +k\left[ D_\mu, D_\nu \right] 
\longrightarrow^{\!\!\!\!\!\!\!\!\!\!\!\!k=2} -D_\mu D_\nu 
- 2{\mathrm{i}}e F_{\mu\nu} .
\ee
However, the choice $k=2$ is only one possibility among the others.
By the trick of such kind one can get the field strength term
with an arbitrary numerical coefficient before $F_{\mu\nu}$. 
Besides, without adding the commutator term $2[P_{\mu},P_{\nu}](=0!)$
to the left side of first equation in (\ref{eq12}) one does not get also from it
by the minimal coupling prescription $P_\mu\to D_\mu$ the $F_{\mu\nu}$
term in eq. (\ref{eq14}). In our theory the field strength term arises
from the $P^2$ term ($UP^2U^{-1} = \pi^2 = D^2 - e F_{\mu\nu}S^{\mu\nu}$)
and eq. (\ref{eq14}) uniquely follows from eq. (\ref{eq12}). Moreover, 
since $\pi_\mu$ and $\pi_\nu$ like $P_\mu$ and $P_\nu$ commute, one gets
by applying the $U$-transformation to eq. (\ref{eq11})  also uniquely
eq. (\ref{eq13}). Due to the dummy 
indices in the free particle equations one cannot apply the
$U$-matrix before transforming the equations into the matrix form. So one
must write for $P_{\mu}P_{\nu}\phi^{\nu}$ term in eq. (\ref{eq11})
$\:P_{\mu}E^{\mu\nu}P_{\nu}\phi\:$ to get 
$\:P_{\mu}E^{\mu\nu}P_{\nu}\phi\to U P_\mu U^{-1} U E^{\mu\nu} U^{-1} U P_{\nu} 
U^{-1} U \phi \to \pi_{\mu} U E^{\mu\nu}
U^{-1} \pi_{\nu}\phi_d \to D_{\mu}E^{\mu\nu}D_{\nu}\phi_d \to 
D_{\mu}D_{\nu}\phi^\nu_d$.
The same result follows also from $P_{\nu}P_{\mu}\phi^{\nu}$. 

For getting unique minimal coupling theory one must depart from the 
first order equations, where the procedure $P_\mu\to D_\mu$ is unambiguous.
However, the Kemmer-Duffin spin-$1$ equation with the minimal coupling
leads to $g=1$, which is in harmony with the old knowledge, that the 
minimal electromagnetic coupling for spin $s$ leads to the gyromagnetic
ratio $g=1/s$ \cite{belinfante,case,fronsdal}.

In the case of the first order equations the ``dynamical'' interaction
is introduced by the modified minimal coupling procedure \cite{saar}
\be
P_{\mu}\to P_{\mu}-eA_\mu-\frac{e}{2(k\cdot P)}k_{\mu}
F_{\rho\sigma}S^{\sigma\rho}=D_{\mu}-\frac{e}{2(k\cdot P)}k_{\mu}
F_{\rho\sigma}S^{\rho\sigma}
\label{eq19}
\ee
The last term in the equation does not give contribution 
to the spin-$0$ and spin-$1/2$ equations. However, in the $s>1/2$ 
cases the added spin-dependent term increases the gyromagnetic ratio 
as compared to the minimal coupling one, leading to the 
$g=2$ value. One can see it more clearly by examining the
spin-dependent terms in the second order equations. Since 
every ``dynamical'' first order equation has the Klein-Gordon
divisor (if such an operator exists for free equation) one can 
always find the corresponding second order equation in the form
given by eq. (\ref{eq10}). Applying, for example, the 
Klein-Gordon divisor to the ``dynamical'' Rarita-Schwinger linear 
spin-$3/2$ equation, one obtains 
\bea
\left[(D^2-m^2-eF_{\rho\sigma}s^{\sigma\rho})g_{\mu\nu}
-2\mathrm{i}eF_{\mu\nu}\right]\psi^\nu_d=0 ,\nonumber\\
\label{eq20}\\
\gamma_{\mu}\psi^\mu_d=0,\nonumber
\eea
where $s^{\sigma\rho}=\frac{\mathrm{i}}{4}[\gamma^\sigma,\gamma^\rho]$ is the
Lorentz spin-$1/2$ generator. Contrary to the minimal coupling case,
where spin-$3/2$ Rarita-Schwinger equation leads to the gyromagnetic
ratio $g=2/3$ \cite{belinfante}, the spin dependent terms in 
eq. (\ref{eq20}) suggest the value $g=2$.

\subsection*{Acknowledgements}
This work has been done in the frame of the project 
no. 0380178S98 financed by the Ministry of Education
of Estonia and partially supported by the Estonian Science Foundation,
grants nos. 3458 and 4510.

\end{document}